\documentclass[aps,twocolumn,groupedaddress,showkeys]{revtex4}

\usepackage{amssymb,amsmath,amsfonts,bm}
\usepackage{graphicx}
\usepackage{color}

\begin{document}

\title{Self-Similarity and Universality
in Rayleigh-Taylor, Boussinesq Turbulence}
\author{Natalia Vladimirova}
\email{nata@flash.uchicago.edu}
\affiliation{ASC Flash Center, The University of Chicago, Chicago, IL 60637},
\altaffiliation{also at CNLS and T-13, Theory Division, Los Alamos National
Laboratory, Los Alamos, NM 87544}

\author{Michael Chertkov}
\email{chertkov@lanl.gov}
\affiliation{CNLS and T-13, Theory Division, Los Alamos National
Laboratory, Los Alamos, NM 87544}

\begin{abstract}
We report and discuss case study simulations of the Rayleigh-Taylor
instability in the Boussinesq, incompressible regime developed to
turbulence. Our main focus is on a statistical analysis of
density and velocity fluctuations inside of the already
developed and growing in size mixing zone. Novel observations
reported in the article concern {\rm self-similarity} of the
velocity and density fluctuations spectra inside of the mixing zone
snapshot, independence of the spectra of the horizontal slice level,
and {\rm universality} showing itself in a virtual independence of
the internal structure of the mixing zone, measured in the re-scaled
spatial units, of the initial interface perturbations.
\end{abstract}


\maketitle

\section{Introduction}

The Rayleigh-Taylor instability occurs when a heavy fluid is being
pushed by a light fluid. Two plane-parallel layers of fluid, colder on
top, are in equilibrium while the slightest perturbation leads to the
denser fluid moving down and the lighter material being displaced
upwards. The early, linear stage of the instability was described by
Rayleigh~\cite{1883Ray} and Taylor~\cite{50Tay}, and summarized
in~\cite{61Cha}. Further development of the instability leads to
enhancement of the mixing and to a gradual increase of the mixing
zone, which is the domain where proportions of heavy in light and
light in heavy are comparable. Dimensional arguments, supported by
large-scale modeling \cite{62DHH,84Sha}, suggest that the half-width
of the mixing zone, $h$, grows quadratically at late time, $h\propto
\alpha A g t^2$, where $A$ is the Atwood number characterizing the
initial density contrast, $g$ is the gravitational acceleration, and
$\alpha$ is a dimensionless coefficient.

The coefficient $\alpha$ was the focus of almost every paper written
on the subject of Rayleigh-Taylor turbulence (RTT) during the last
fifty years.  The first attempts to look inside the mixing zone were
initiated only in late 1990s~\cite{94Y,94LRY,99DLY}, due to advances
in experimental and numerical techniques. The results of many studies
and the controversies surrounding the $\alpha$-coefficient were
recently summarized in the review combining and analyzing the majority
of existing $\alpha$-testing simulations and
experiments~\cite{04Dim}. In this article we also discuss the
developed regime of RT turbulence.  Our main focus is on the analysis
of the internal structure of the mixing zone, and we trace the
$\alpha$-coefficient only for validation purposes.

Our analysis of the mixing zone develops and extends previous
experimental ~\cite{99DLY,02WA,04RA} and numerical
~\cite{99DLY,01YTDR,01CD,04RC,04CCM,06CC,06DT} observations on the
subject, and it is also guided by phenomenological considerations
discussed in~\cite{03Che}. The essence of the phenomenology, which
utilizes the classical Kolmogorov-41 approach~\cite{41Kol}, can be
summarized in the following statements: (i) The mixing zone width,
$h$, and the energy containing scale, $R_0$, are well separated from
the viscous, $\eta$, and diffusive, $r_d$, scales. In the inertial
range, realized within the asymptotically large range bounded by
$R_0/\eta$ from above/below, turbulence is adjusted adiabatically to
the large-scale buoyancy-controlled dynamics. (ii) In three
dimensions, the velocity fluctuations at smaller scales are
asymptotically decoupled from weaker buoyancy effects~\footnote{In two
dimensions RTT is markedly different from its three dimensional
counterpart: buoyancy and inertia effects are in balance, resulting in
the so-called Bolgiano-Obukhov scaling regime. The $2d$
phenomenological prediction of \cite{03Che} was numerically confirmed
in~\cite{06CMV}.}. (iii) Typical values of velocity and density
fluctuations scale the same way as in the stationary, homogeneous
Kolmogorov turbulence, $\delta v_r\sim (\epsilon r)^{1/3}$, and
$\delta \rho_r\sim \epsilon_\rho^{1/2}\epsilon^{-1/6} r^{1/3}$, where
the energy Kolmogorov flux, $\epsilon$, increases with time while the
density fluctuations flux, $\epsilon_\rho$, remains constant,
according to the buoyancy prescribed large scale dynamics.  All of
these three theses of the phenomenology are consistent with available
experimental~\cite{99DLY,02WA} and
numerical~\cite{01YTDR,01CD,05ZWRDB,06CC} observations of the velocity
and density spectra.  One particularly important consequence of the
phenomenology, the decrease of the viscous and dissipative scales, was
also predicted in~\cite{04RC} and numerically confirmed
in~\cite{04RC,06CC}.

In spite of its relative success in explaining RTT,  the
phenomenology \cite{03Che} is, obviously, not free from
deficiencies. First, the asymptotic, large time character of the
theory turns into a handicap in explaining numerical and
experimental data, taken at finite,  and actually modest, times.
Second,  the phenomenology treats all $z$-slices within the mixing
zone equally. Third,  the phenomenology does not differentiate
between the mixing zone width, $h$, and the energy containing scale,
$R_0$, for the turbulent fluctuations.

\begin{figure*}
\begin{center}
\includegraphics[width=\textwidth]{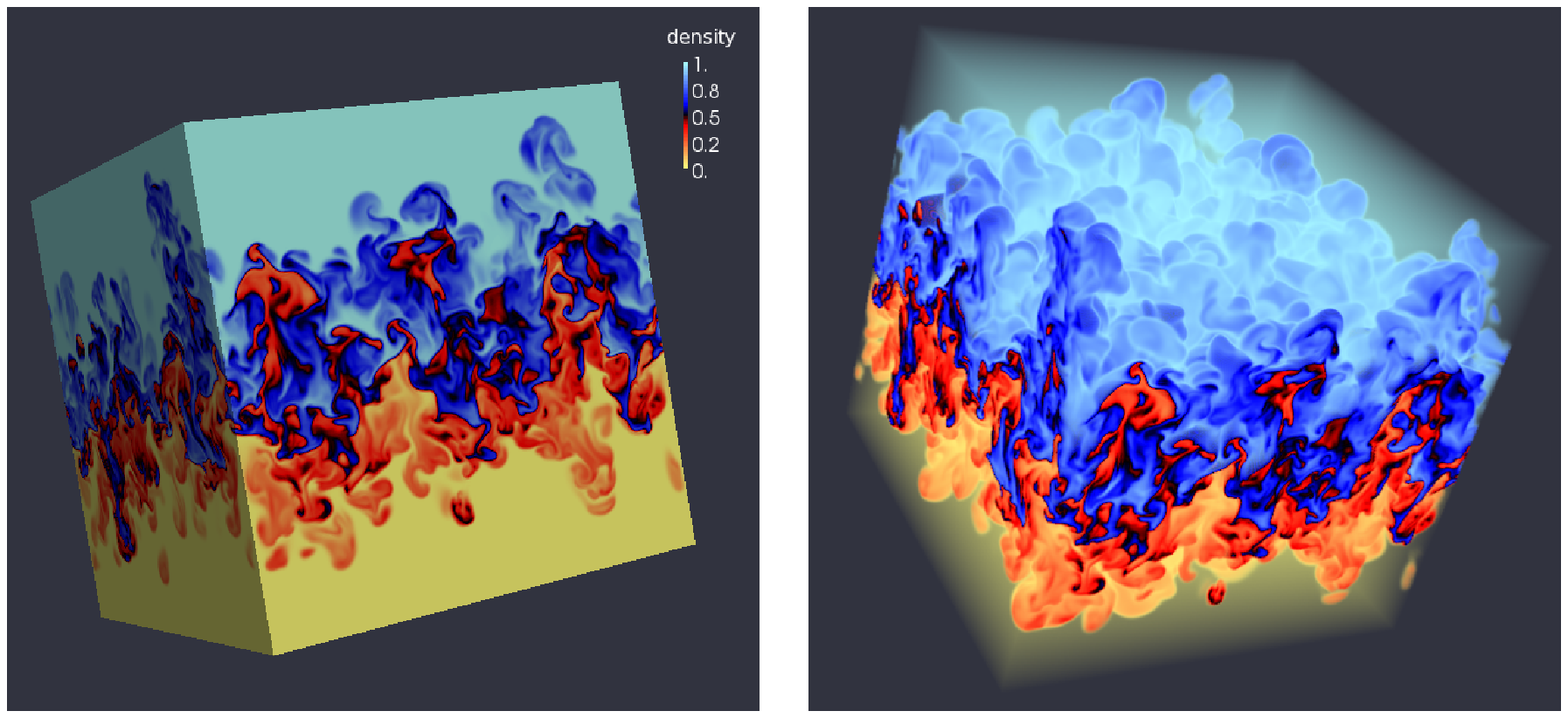}
\end{center}
\caption{
   Density at $t=128$ in the simulation with narrow initial 
   spectrum in $960^3$ domain.}
\label{fig_cubes}
\end{figure*}

Improving the phenomenology from within itself, or by some
complementary theoretical means, does not seem feasible, and one needs
to rely on resolving these questions/uncertainties through experiments
and simulations. This article reports a step in this direction.  Here
we raise and give partial answers, based on simulations, to the
following subset of key questions concerning the internal structure of
the RTT mixing zone:
\begin{itemize}
\item
   Analyzing the evolution of $h$, $R_0$, $r_d$ and $\eta$ with time one
   often observes a non-universal, simulation/experiment specific
   behavior, especially at transient, so-called early self-similarity,
   times~\cite{04RC}.
   Will the relative dependence of scales be a more reliable indicator of
   a universal behavior than the time-dependence of the
   scales?
\item
   How does the energy containing scale, $R_0$, compare with the width
   of the mixing layer, $h$? This question was already addressed in
   \cite{04RC}. Here, we will elaborate on this point.
\item
   How different are the turbulent spectra at
   different vertical positions in the mixing zone within a given time snapshot?
\item
   How different are the scales and spectra corresponding to
   qualitatively different initial perturbations?
\end{itemize}

The material in this article is organized as follows. We start by
describing our simulations, we then proceed to the definitions and
subsequently the observations of the various spatial scales
characterizing snapshots of the mixing zone. Finally we discuss
self-similarity and universality of the emerging spatio-temporal
picture of the RTT.  We conclude by answering the questions posed above.

\section{Description of simulations}

\begin{figure*}
\begin{center}
\includegraphics[width=\textwidth]{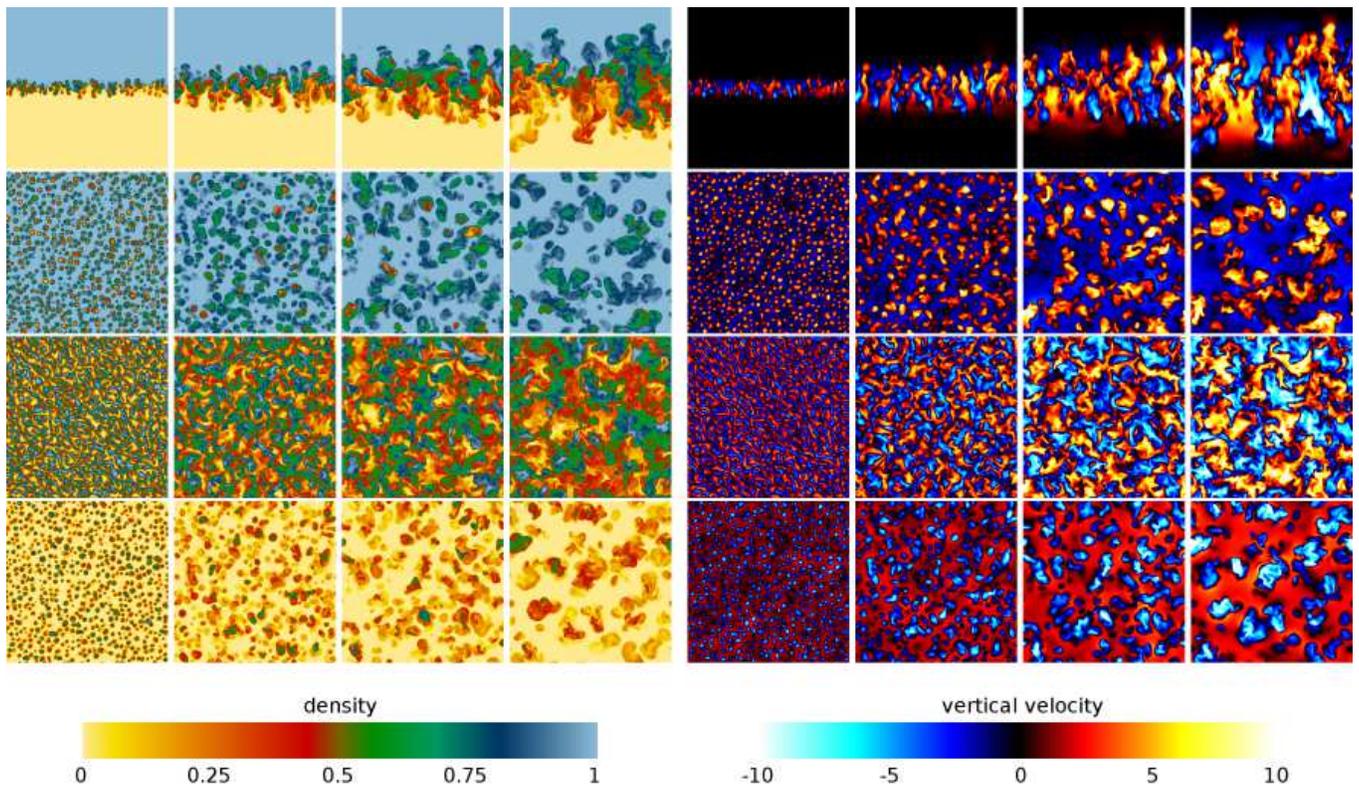}
\end{center}
\caption{
   Slices of density and vertical velocity at times $t=32$, 64, 96, 128 (left to
   right) in simulation with narrow initial spectrum in $960^3$ domain;
  from top to bottom, the images correspond to vertical slice
   at $y=480$ and horizontal slices at $z=+0.75h$, $0$, $-0.75h$.}
\label{fig_color}
\end{figure*}

We consider 3D incompressible, miscible Rayleigh-Taylor flow in the
Boussinesq regime,
\begin{eqnarray}
&& \partial_t {\bm v}+({\bm v}\nabla){\bm v}+\nabla p-\nu\Delta{\bm
v}= A{\bm g}c,\quad\nabla {\bm v}=0,\label{NS}\\ &&
\partial_t c+({\bm v}\nabla)c=\kappa\Delta c,\label{bm_v}
\end{eqnarray}
where $\kappa$ and $\nu$ are the diffusion and viscosity coefficients,
while $c=(\rho-\rho_{\rm min})/(\rho_{\rm max}-\rho_{\rm min})$ is the
normalized density.  The Boussinesq approximation for gravity
corresponds to fluids with small density contrast, $A\ll 1$, where
$A=(\rho_{\rm max}-\rho_{\rm min})/(\rho_{\rm max}+\rho_{\rm min})$ is
the Atwood number.  Here, we restrict ourself to the case of $\kappa =
\nu$.

We solve equations~(\ref{NS}-\ref{bm_v}) using the spectral element
code of Fischer et al~\cite{code} designed specifically for DNS of
Boussinesq fluids.  The equations are solved in the nondimensional
units,
\[
   [l] = (2Ag)^{-\frac{1}{3}} \nu^{\frac{2}{3}}, \hspace{15mm}
   [t] = (2Ag)^{-\frac{2}{3}} \nu^{\frac{1}{3}};
\]
the results are presented in the same units.  This choice of units is
based solely on the dimensional parameters in
Eqs.~(\ref{NS}-\ref{bm_v}) thus reflecting free boundary conditions
(absence of any wall constraints). The critical wavelength and the
wavelength of the linearly most unstable mode are constants in these
units.

The boundary conditions are periodic in the horizontal directions
and no-slip in vertical direction.  The initial conditions include a
quiescent velocity and a slightly perturbed interface between the
layers, $c(t=0;z)=-\theta(z+\delta(x,y))$, where
$\theta(z)=\frac{1}{2}\left[ 1 - \tanh(0.4z) \right]$ is the
function describing the density profile across the interface and
$\delta(x,y)$ is the perturbation.

We use spectral elements of size $30^3$ with 12 collocation points in
each direction. This is equivalent to the spectral resolution with
the spacing between points $\Delta = 3$.
The size of our largest computational domain is $1920\times
1920\times 1440$ physical units, or $768\times 768\times 576$
collocation points.  We stop our simulation at, $t=128$, when the
width of the mixing layer reaches the domain size.  At the end of
simulation, the Reynolds number reaches $\Re = 7500$ to $\Re=13000$
depending on the initial conditions, where $\Re =
\frac{4 h \dot{h}}{\nu}$.  For comparison, the largest Rayleigh-Taylor
simulation to date~\cite{06CC} was performed in a $3096^3$ domain at
resolution $\Delta=1$ and reaching time $t=248$ and $\Re = 30000$.
Our relatively coarse resolution might raise concerns, especially in
diagnostics of small structures.  Nevertheless, all our results,
including the spectra and the estimates for microscale $\eta$, are
in a very good agreement with~\cite{06CC}, as well as with our
finer-resolved (but smaller) simulation with $\Delta=1$.

In all cases studied, the (initial) fastest growing mode is located at
$\lambda \approx 24$. Most of the presented results were obtained in
the simulations in the domain of $1920\times 1920\times 1440$ physical
units with initial perturbation in the form of a narrow initial
spectrum, with modes $36 \le n \le 96$ and spectral index~$0$.  (Here
the spectral index refers to the exponent of the wavenumber, as
in~\cite{05RDA}, and describes the shapes of the spectra.)  To
investigate the influence of initial condition we also performed
additional simulations in smaller domain of size $960^3$ (in physical
units) with: (1) a narrower initial spectrum, with modes $18\le n \le
48$ and spectral index~0 (Figure~\ref{fig_cubes}); and (2) a broader
initial spectrum, with modes $3\le n \le 96$ and spectral index~$-1$.
The two regimes were identified in previous studies as giving
distinctly different $h(t)$ at transient times~\cite{05RDA}. According
to~\cite{05RDA}, the first system develops a mode-merging regime and
exhibits scalings with universal $\alpha$, while the second system
develops in the regime of mode-competition, with $\alpha$ depending on
the amplitude of the initial perturbation. We observed that, in spite
of the early stage differences, the additional simulations gave the
same results in the turbulent (advanced time) regime as the main set.

One important focus of our simulations/analysis is on resolving the
vertical inhomogeneity of the mixing zone (Figure~\ref{fig_color}).
To achieve this goal we differentiate vertical slices within a given
snapshot, thus calculating various characteristics of the mixing
zone such as the energy containing scale, the energy spectra and the
viscous scale. We collect statistics within a given slice $z$, e.g.
contrasting results for the mixing zone center and its periphery.

\section{Scales of Rayleigh-Taylor Turbulence}

\subsection{Mixing zone width}

\begin{figure}
\includegraphics[width=0.47\textwidth]{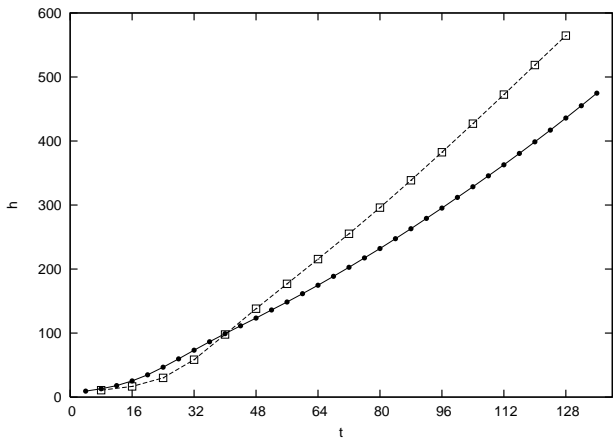}
\caption{
   Mixing widths $h$ in simulations with narrow
   (solid line) and wide (dashed line) initial spectra. }
\label{fig_mixing_width}
\end{figure}

The mixing zone width is the standard characteristic used in
the $\alpha$-studies \cite{04Dim}. According to the most recent analysis
\cite{04RC,06CC}, the mixing zone width obeys the scaling, $\sqrt{h}
= \sqrt{h_0} + t \sqrt{\alpha Ag}$, where $h_0$ is an
initial-conditions-dependent constant.  In the simulation with
narrow initial spectrum, we reproduce this scaling relatively well;
in the faster-developing simulation with a broad initial spectrum
the scaling is affected by the finite domain size
(Figure~\ref{fig_mixing_width}).

\begin{figure*}
\includegraphics[width=\textwidth]{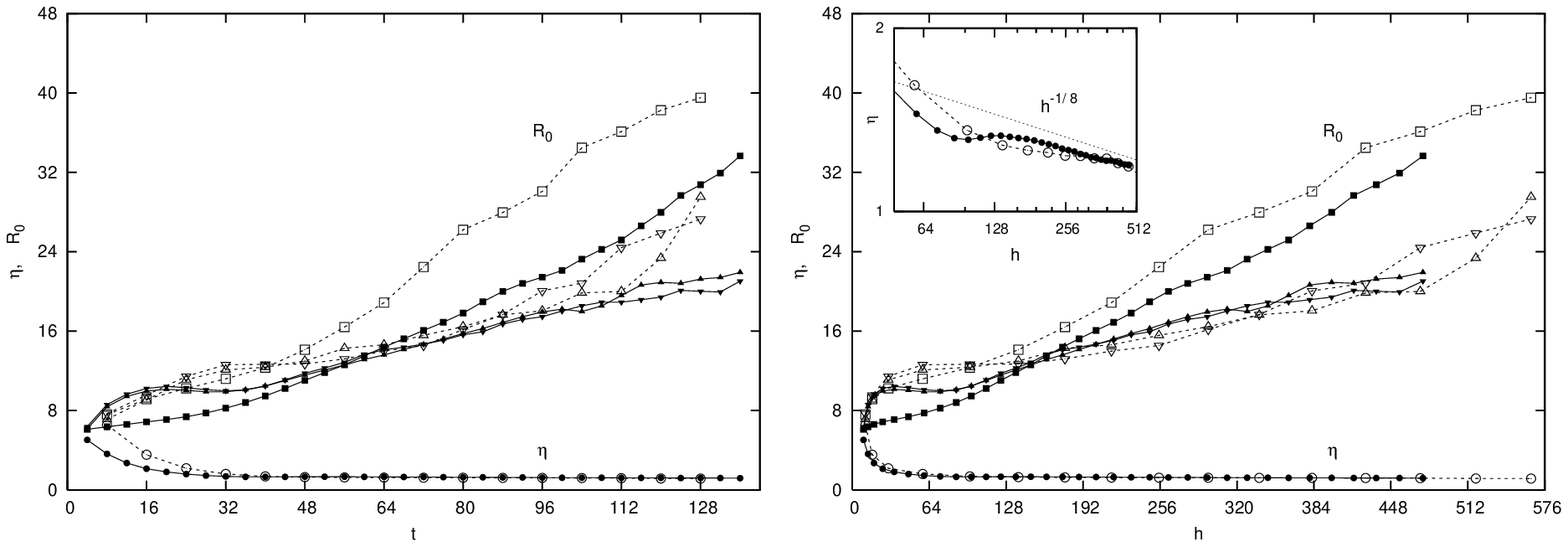}
\caption{
   The energy containing scale (correlation length) and the viscous
   scale in the middle of mixing layer in the simulations with narrow
   (solid line) and broad (dashed line) initial spectra.  Squares and
   triangles correspond to scales computed using vertical and
   horizontal components of the velocity, respectively.  }
\label{fig_scales}
\end{figure*}

\begin{figure*}
\includegraphics[width=\textwidth]{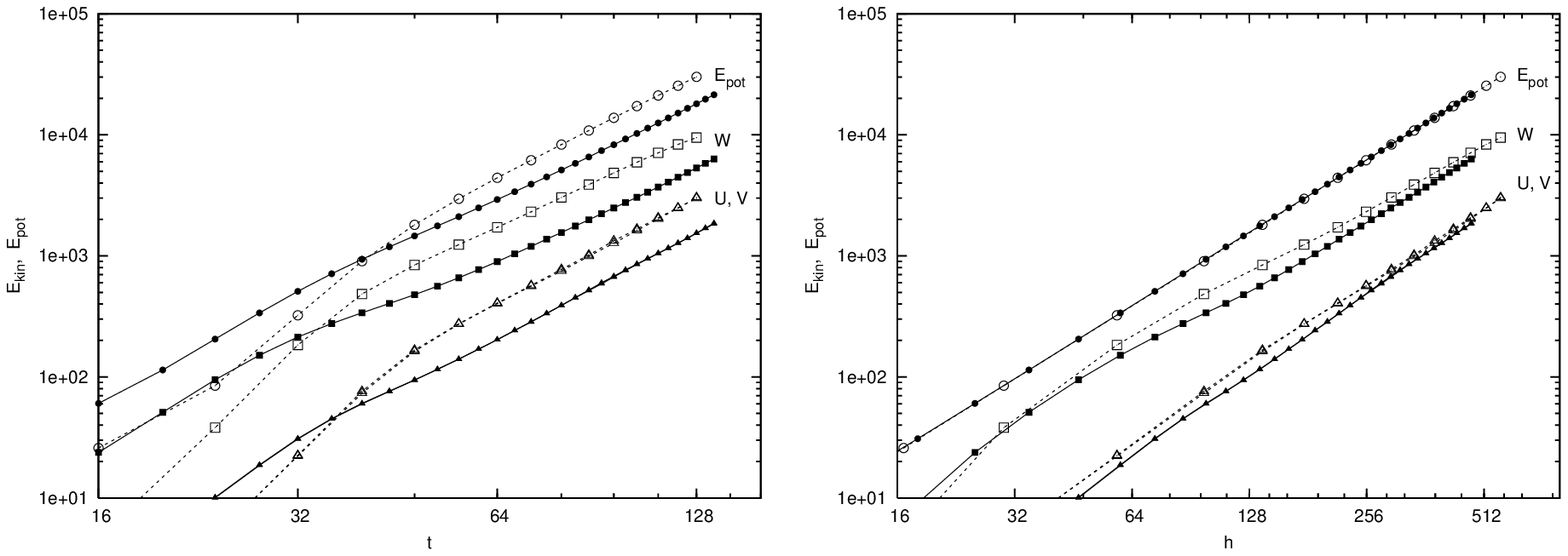}
\caption{
   Released potential energy and components of the kinetic energy
   integrated across the mixing layer in simulations with narrow
   (solid line) and broad (dashed line) initial spectra.  }
\label{fig_energies}
\end{figure*}

The value of $\alpha$ depends on the definition of the mixing zone
width.  Following~\cite{90AS,99DLY,06CC} we consider the definition
based on the mixing function, $M(c) = 4c(1-c)$,
\begin{equation}
   h = \int M(\bar{c}) dz,
   \label{eq_hdef}
\end{equation}
where the overbar denotes averaging over the horizontal plane.  We
prefer integral definitions of the mixing zone width over the common
definitions based on the values of $\bar{c}$ (for example the
half-distance $H$ between two heights where $\bar{c}=0.01$ and
$\bar{c}=0.99$) simply due to the fact that integral quantities are
less sensitive to the profile of $\bar{c}(z)$ at the edges of the
mixing layer, and consequently the quantities are less sensitive to
the size of the computational domain. In the established self-similar
regime unrestricted by domain boundaries, two definitions of the
mixing zone width are actually within an $O(1)$ systematic factor of
each other.  The value of the coefficient $\alpha$, determined from
the slopes of the curves shown in Figure~\ref{fig_mixing_width} at
$t>60$ are $\alpha=0.029$ for the narrow initial spectrum and
$\alpha=0.040$ for the broad initial spectrum. The obtained values of
$\alpha$ are in a good agreement with other simulations (see reviews
in~\cite{04Dim,04RA}).  Experimental values are higher, $\alpha\approx
0.5$-$0.7$, which is usually attributed to the presence of longer
wavelengths in the initial spectra, and our simulation with broader
initial spectrum follows the same tendency.  As we show below, in
spite of the difference in $\alpha$, the two systems are very similar.
Further broadening of the initial spectrum would require larger a and
more expensive simulation, while we do not expect the results to be
significantly different.

An important thesis of the phenomenology \cite{03Che} is that the
internal structure of the mixing zone senses the overall time scale
only adiabatically through slowly evolving large scale
characteristics, of which the mixing zone width, $h$, is the benchmark one.
Therefore, our intention is to separate the ``large
scale'' question of the overall time dependence of the mixing zone
width from the set of focused ``small scale'' questions about
internal structure of the mixing zone. To achieve this goal, we
track the dependence of the various internal characteristics of the
mixing zone (see below) on the mixing zone width.

\subsection{Energy-containing scale}

The energy-containing scale represents the size of a typical turbulent
eddy which, intuitively, corresponds to the size of the large scale
vortices seen in the mixing zone snapshot, e.g. shown in
Figure~\ref{fig_color}. Formally, it is convenient to define this
scale, $R_0$, as the correlation length of the normalized two-point
pair correlation function of velocity, $f(R)=\langle
v_i(r)v_i(r+R)\rangle/\langle v_i^2 \rangle$, where $v_i$ is one
spatial component of the velocity vector, ${\bm v}$. We estimate $R_0$
as a half width of the correlation function, $f(0)/f(R_0)=2$. Defined
this way, $R_0$ is consistent (up to some $\pi$-dependent constant)
with the wavelength (inverse of the wave vector) where the turbulent
energy spectra achieves its maximum. See e.g.
Figure~\ref{fig_spectra}.

\begin{figure}
\includegraphics[width=0.47\textwidth]{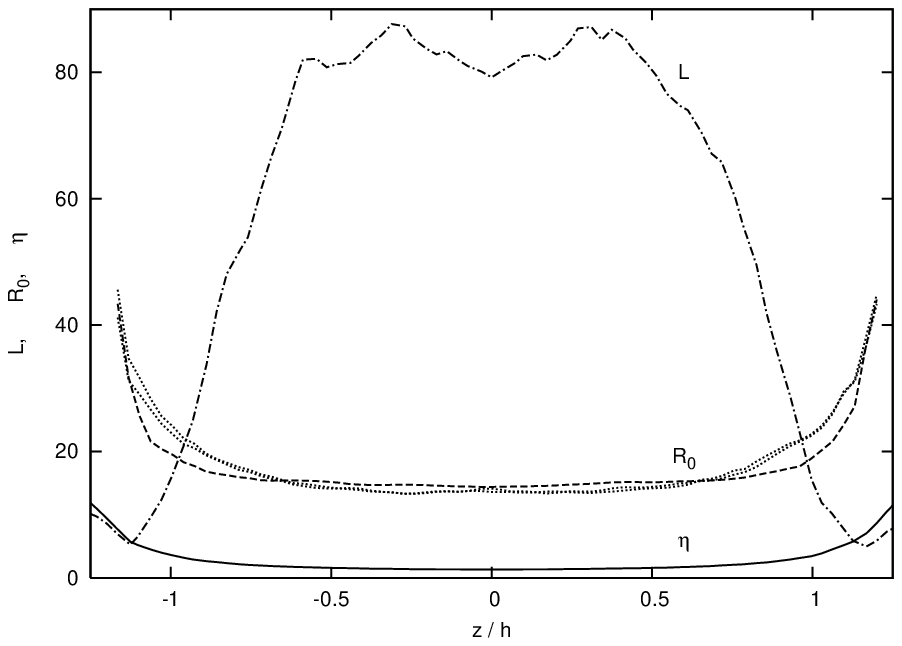}
\caption{
   $R_0$ -- energy containing scale based on the two-point correlation function,
   $L$ -- energy containing scale based on the single point measurements (see discussion in the text for details), and $\eta$ -- the viscous scale, all plotted at time $t=64$ in simulations with narrow initial spectrum as  functions of
   the distance from the middle of the mixing layer.
  Dashed and dotted lines correspond to scale $R_0$ computed using vertical
   and horizontal components of velocity respectively.}
\label{fig_scales_vsZ}
\end{figure}

\begin{figure*}
\includegraphics[width=\textwidth]{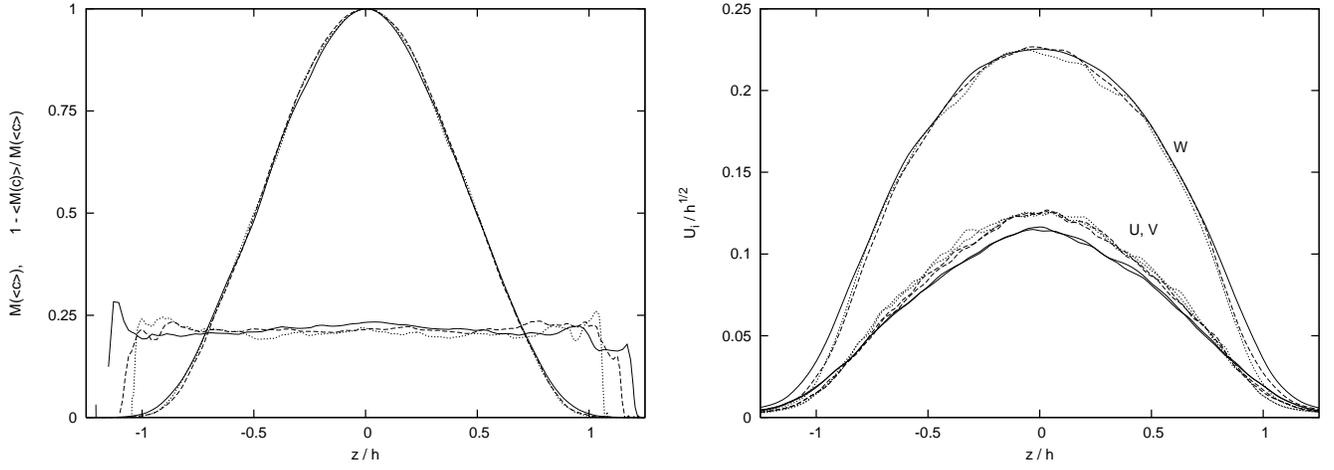}
\caption{
   Mixing function (left) and velocities (right) vs distance from the
   center of the mixing layer.  The quantities are shown at times
   $t=64$ (solid line), $t=96$ (dashed line), and $t=128$ (dotted
   line) for narrow initial spectrum.} \label{fig_velotemp_vsZ}
\end{figure*}

\begin{figure*}
\includegraphics[width=\textwidth]{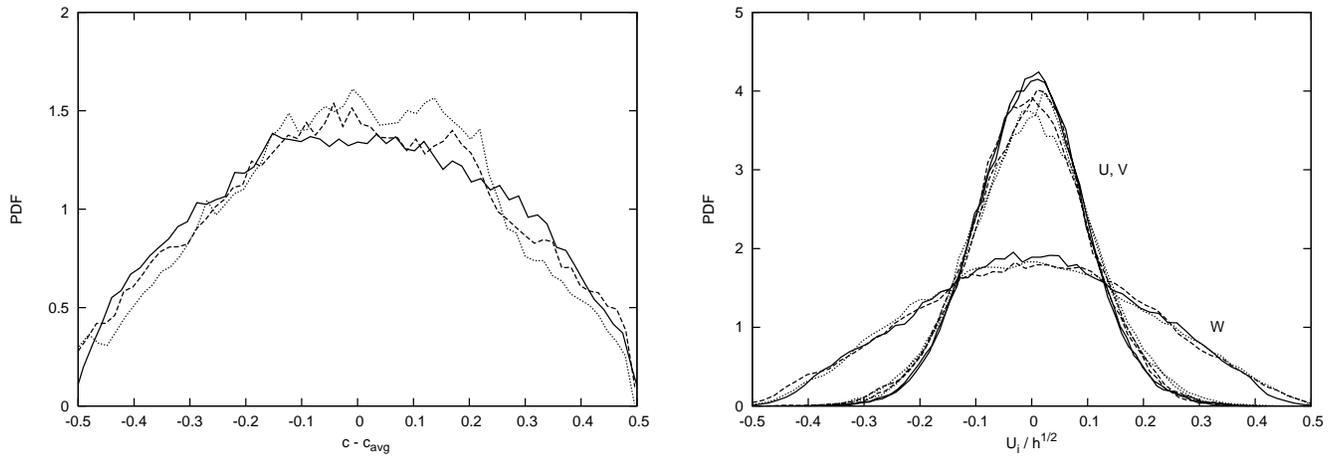}
\caption{
   PDF of density (left) and velocities (right) in
   the center of the mixing layer at times $t=64$ (solid
   line), $t=96$ (dashed line), and $t=128$ (dotted line)
   for narrow initial spectrum.} \label{fig_pdf_z0}
\end{figure*}

\begin{figure*}
\includegraphics[width=\textwidth]{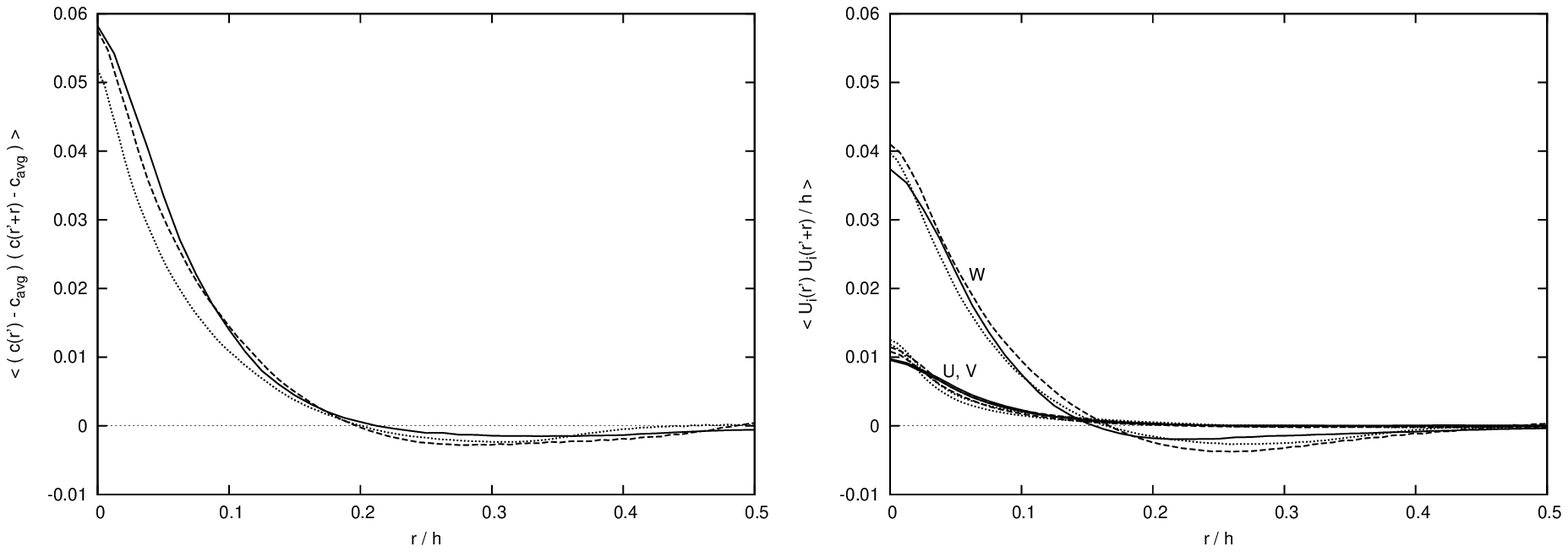}
\caption{
   Correlation functions of density (left) and velocities (right) in
   the center of the mixing layer at times $t=64$ (solid
   line), $t=96$ (dashed line), and $t=128$ (dotted line)
   for narrow initial spectrum.} \label{fig_pcf_z0}
\end{figure*}

In the developed regime, the correlation length taken at the center
of the mixing zone grows linearly with the mixing layer width, $R_0
\approx h/30 + 7$ and $R_0 \approx h/17 + 6$ for correlations
between horizontal and vertical velocities respectively for both
types of initial perturbation (Figure \ref{fig_scales}). In the
limit of large $R_0$, this suggests significant separation between
the two scales, $h:R_0 = 17:1$ or more.

Ristorcelli and Clark~\cite{04RC} introduced their version of the
energy containing scale as $L = v_z^3 / \epsilon$, and found this
scale to be of the order of the width of the mixing layer, $L
\approx 0.4 h$. Note that defining $L$ requires a single point
measurement, while $R_0$ characterizes the two-point correlations.
This single-point nature of $L$ makes it the preferred large scale
characteristic in the engineering closure modeling. Our simulations
show that $L$ is significantly larger than $R_0$, where both $L$ and
$R_0$ scale linearly with $h$: $L \approx 7-15 R_0 \approx 0.5 h$.
The scale separation of $L$ and $R_0$ may be viewed as a very
favorable fact for the engineering modeling of the RTT, thus
suggesting a numerical justification for the closure schemes, e.g.
of the type discussed in \cite{04RC}.

\subsection{Viscous and dissipation scales}

In our simulations $\kappa=\nu$, thus the viscous scale and the dissipation
scale are equal to each other.   (This simplification reflects our desire to focus on the larger scale physics while keeping the resolution domain sufficiently large.) We estimate the viscous scale in the middle of the mixing layer
($z=0$) directly as $\eta\sim (\nu^3/\epsilon)^{1/4} \sim (\nu^2/
\langle 5 (\nabla v)^2 \rangle )^{1/4}$. (The ``$5$''-factor here is
an artifact of an old tradition, see e.g. \cite{72TL}.)
In magnitude, the viscous scale agrees with~\cite{06CC} and with the
respective phenomenology estimate ~\cite{04RC,03Che},
$\eta\sim\left( (\nu/v)^3 h\right)^{1/4}$. The viscous scale
decreases slowly with time (see Figure~\ref{fig_scales}).  However,
our data are too noisy to claim anything more than rough consistency
with the $h^{-1/8}$ predicted in \cite{04RC,03Che} and observed in
\cite{04RC,06CC}.

\subsection{Relative dependence of scales}

In view of our focus on the internal structure of the mixing zone,
we choose to study the relative dependence of the relevant scales.
Thus, Figure~\ref{fig_scales_vsZ} shows dependence of the
energy-containing and viscous scales on the mixing zone width.

Analyzing simulations of RTT corresponding to different initial
perturbations, we confirm the earlier observed~\cite{01CD,04RC}
sensitivity of time-evolution of the mixing zone width, scales
$\eta$ and $R_0$, and the integral quantities on initial conditions.
(See left panels in Figures~\ref{fig_scales} and \ref{fig_energies}.)
However, we also find that the same quantities re-plotted as
functions of $h$ look very much alike (Figures~\ref{fig_scales} and
\ref{fig_energies}, right panels). Therefore, one conclusion we
draw here is that the relative scale representation is actually a
better universal indicator of turbulence within the mixing zone.

\section{Structure of the mixing layer}

\subsection{Self-similarity}

Figure~\ref{fig_velotemp_vsZ} shows dependencies of the mixing function and of the
RMS-averaged velocities across the mixing layer on the height, $z$.  In magnitude, the velocities are $O(h^{1/2})$,
as suggested by the $h^2$-dependence of the total kinetic energy: $E
\sim v^2 h$ (See Figure~\ref{fig_energies}). The curves taken at
three different times are almost indistinguishable from each other.
This suggests that the mixing zone, viewed from the
large-scale perspectives, is self-similar.

Self-similarity of the averaged density and the averaged velocities
was observed in~\cite{01CD,01YTDR,04CCM,04RA,04RC,06DT}. The
self-similarity itself does not define the specific form of the
$z$-averaged profiles, only suggesting that these profiles are smooth
functions of $z/h$. In particular, self-similarity is in principle
consistent with specific parabolic predictions for the size of the
dominant eddy and total kinetic energy contained in the mixing layer
\cite{96SA,06DT}, $K(z)=K_0(1-z^2/h^2)$ and $L(z)=L_0\sqrt{1 -
z^2/h^2}$, where $K_0$ and $L_0$ are respective characteristics
measured in the middle of the mixing layer.  Notice, however, that our
simulation results, shown in Fig.~\ref{fig_scales_vsZ}, suggest a
much flatter profile for kinetic energy than the parabolic one.

We notice that
when illustrating self-similarity (see for instance~\cite{04RC}), it
is common to rescale the mean profiles of different quantities using the
values at the center of the mixing layer and plot these profiles as
function of $z/h$.  Here, we propose to use $h$-based scalings not
only for the $z$-coordinate but also for the discussed quantity.  Thus, in
Figure~\ref{fig_velotemp_vsZ} we rescale velocities with $\sqrt{h}$.

In addition to the mixing function dependence on $z$,
Figure~\ref{fig_velotemp_vsZ} shows $1-\theta(z) = 1 -
\overline{M(c)}) / M(\overline{c})$ to be almost constant
inside of the mixing layer, with a sharp drop-off near the edges.
Profiles of $\theta(x)$, sometimes called the molecular mixing fraction,
were obtained experimentally and numerically
in~\cite{91Youngs,94LRY,99DLY,04Dim,04RA}, and most of the
observations agree on the fact that at the later stages of the RT
instability $\theta(x)$ remains constant across the mixing layer at
approximately 0.75-0.8.  This suggests that a
$\theta$-based definition of the mixing zone width can be
advantageous because it generates a more robust scaling.

The $h$-scaling and self-similarity of the mean profiles are also
observed in the probability distribution functions (PDFs) as well as
in the correlation functions for density and velocities computed at
$z=0$ (see Figures~\ref{fig_pdf_z0} and~\ref{fig_pcf_z0}).  
The self-similarity is observed at sufficiently large times, $t>64$, 
but is lacking at the earlier times (not shown in the
Figures). The dynamics one sees at the earlier times for the PDF
points to transition from a single peak curve to two peaks and to a
single peak again.  The explanation for this phenomenon is as follows. The
initial, single peak distribution is dependent on the initial
perturbation of the originally sharp interface.  The transformation from
one to two peaks corresponds to transition to the non-linear regime of the
RT instability, associated with the secondary Kelvin-Helmholtz type shear
instability and the formation of RT mushrooms.  The transition from two
peaks to one corresponds to the destruction of RT mushrooms and
formation of the turbulent mixed zone. Notice that the emergence of the
earlier time transitions is consistent with results reported 
in~\cite{99DLY,04CCM,04RA} for the PDF of density, where, probably, the
asymptotic self-similar regime was not reached yet.

\subsection{Dependence of scales on $z$-location}
\begin{figure*}
\includegraphics[width=\textwidth]{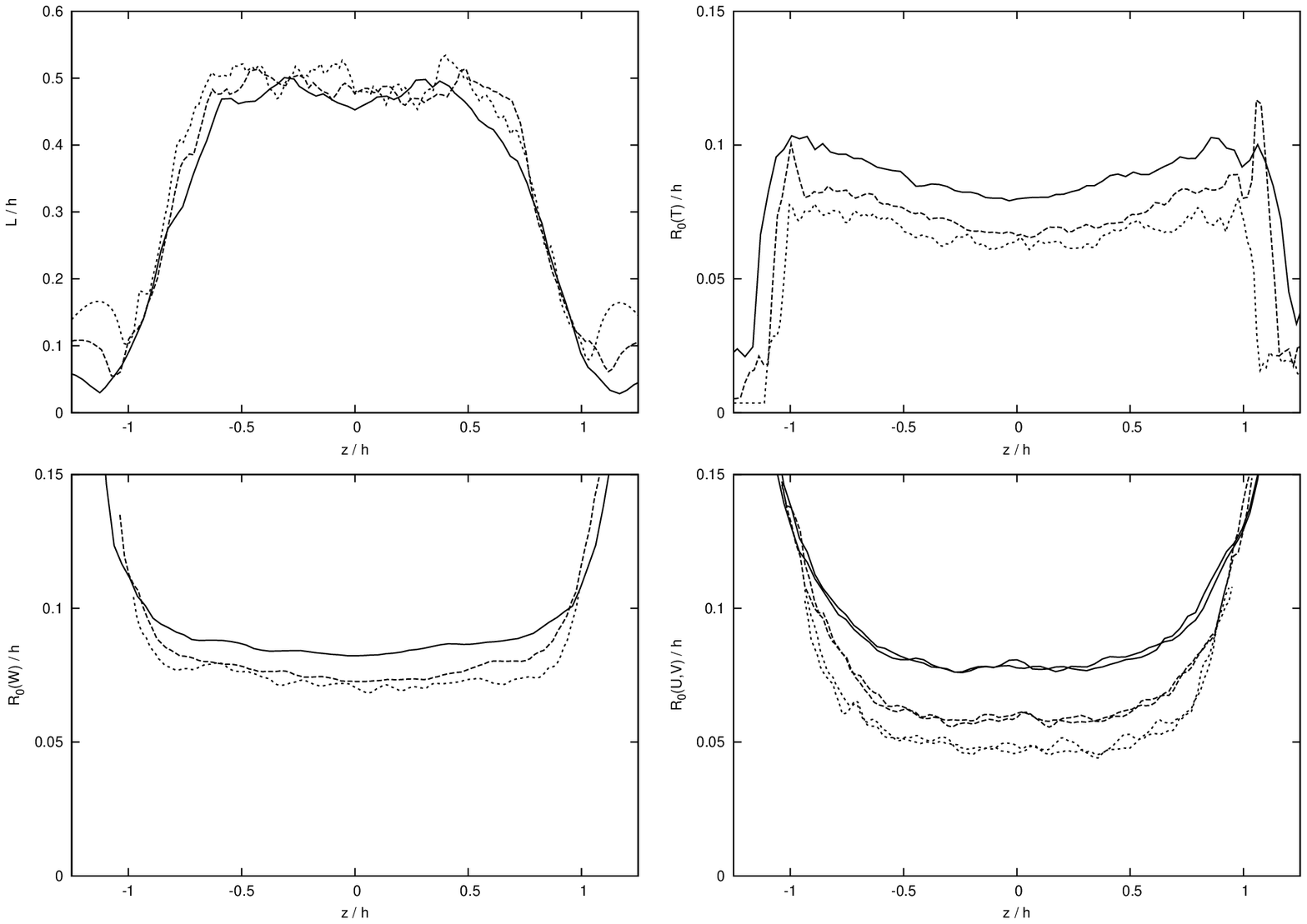}
\caption{
   The energy containing scale based on the single point measurements,
   $L$, and based on the two-point correlations of density and
   velocities, $R_0$, as function of the distance from the center of
   the mixing layer at times $t=64$ (solid line), $t=96$ (dashed
   line), and $t=128$ (dotted line) for narrow initial spectrum.
}
   \label{fig_scales_vsZ_2}
\end{figure*}

A discussion of scales is rare in the existing literature, not to
mention a discussion of how the scales vary across the mixing zone.
This is partly due to difficulties with experimental diagnostics.
Notable exceptions are~\cite{04RC} and~\cite{06DT}, which discuss the
energy containing scale within simulations and one-dimensional
modeling, respectively.  The model in~\cite{06DT} predicts a parabolic
profile for the energy containing scale, and this appears consistent
with earlier time observations in~\cite{04RC} and in our setting.  At
later stages the profile in~\cite{04RC} changes to a much flatter one,
which also agrees with our estimates of the energy containing scale based
(as in~\cite{04RC}) on single point measurements.  Moreover, we show
that later in time these profiles stabilize in a nontrivial
self-similar solution (Figure~\ref{fig_scales_vsZ_2}).

Our measurements, based on the two-point correlation functions of density
and velocities as well as estimates of the dissipation scale
$\eta$, also exhibit monotonic dependence on $z/h$.  As shown in
Figure~\ref{fig_scales_vsZ}, both $R_0$ and $\eta$ vary
insignificantly across the mixing layer with a slight increase towards
the edges.  Among the scales we measured, $R_0$ is the only scale
which the self-similarity in time is still questionable
(Figure~\ref{fig_scales_vsZ_2}).   It might be related to
a resolution-related systematic error in obtaining the correlation
functions near zero (Figure~\ref{fig_pcf_z0}).

Comparing the respective curves for different initial spectra, we find
that the dependence of $R_0(z/h)$ and $\eta(z/h)$ on initial perturbation
is weak.

\subsection{Energy spectra}

\begin{figure*}
\includegraphics[width=\textwidth]{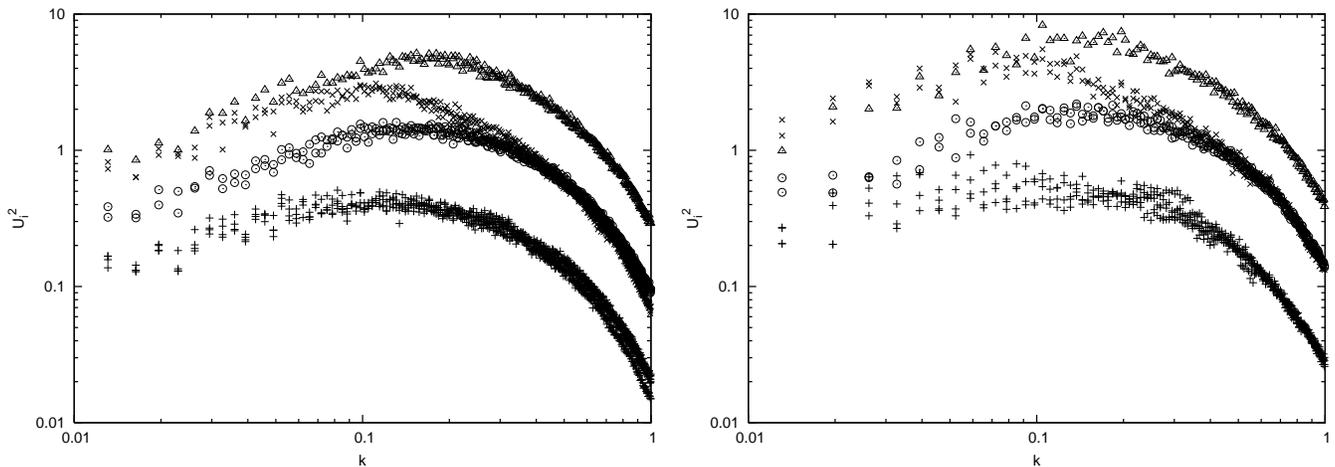}
\caption{
   Energy spectra at time $h \approx 180$ in horizontal planes $z=0$ (circles -
   horizontal velocities, triangles - vertical velocities) and $z \pm
   0.75 h$ ($+$ horizontal velocities, $\times$ vertical velocities)
   in simulations with narrow (left) and broad (right) initial spectra.
   }
\label{fig_spectra}
\end{figure*}

As expected the level of fluctuations grows with time, resulting in a
monotonic shift of the turbulence spectra maxima towards larger
wavelengths. The energy-containing wavelength, $\lambda_0$,
(corresponding to the maximum) is in agreement with the correlation
radius, $R_0$. This applies to various spectra. For example, the
maximum of the $W^2$-spectrum obtained at $z=0$ and $t=64$, and shown
in Figure~\ref{fig_spectra} (right), is located at $k\approx 0.15$,
i.e. $\lambda \approx 42$. The correlation radius at this time is $R_0
\approx 14 \sim
\lambda /\pi\sqrt{2}$.  At time $t=124$ the spectrum maximum shifts
to $k=0.07$ ($\lambda \approx 90$) while $R_0 \approx 30$.

Extracting the Kolmogorov scaling for velocity fluctuations at
different wavelengths as well as the largest wavelength cutoff
(corresponding to the inverse of the viscous scale) from the spectral
data (e.g. Figure~\ref{fig_spectra}) is problematic due to the lack of
spatial resolution. In this regard our simulations, as well as many
others, e.g. \cite{99DLY,01CD,04Dim,04CCM}, lack the power of the
simulations performed by Cook and Cabot \cite{04CCM,06CC}: the latter
provide the only reliable confirmation (so far) of the expected
Kolmogorov features of the spectra. (These record simulations are an
LES run~\cite{04CCM} and a DNS run with $3096^3$ points
resolution~\cite{06CC}.)  Based on our simulation results, we can only
state that the range of scales compatible with the Kolmogorov scaling
grows with time and that the viscous scale decreases with time in
accordance with predictions of
\cite{04RC,03Che}.

\section{Conclusions}

We conclude by presenting a question-and-answer style summary for the
observations made in the article.

\begin{itemize}
\item
{\em Does the relative dependence of the characteristic scales
constitute a better indicator of self-similarity within the RTT than the
dependence of the individual scales on time?} We found
that the energy-containing scale, $R_0$, and the viscous scale,
$\eta$, exhibit monotonic evolution with $h$. At transient times
both $R_0$ and $\eta$ demonstrate much clearer scaling with $h$ than
with the observation time $t$.
\item
{\em How does the energy containing scale, $R_0$, compare with $h$?}
We found that at late time, the ratio of $R_0$ to $h$ taken at the
center of the mixing zone is $\approx 1:20$ and fluctuates little with time.
\item
{\em Do the turbulent spectra vary with vertical position of the
mixing zone?}  Analyzing spatial correlations for a given time
snapshot, we did not observe any qualitatively new features of the
turbulence spectra with transition from the vertically central slice
to an off-centered one within the mixing zone.
\item
{\em How different are the scales and the spectra corresponding to
qualitatively different initial perturbations?}  We found that the
dependencies of $R_0$ and $\eta$, as functions of $z/h$, on the
initial perturbations are weak.
\end{itemize}

Our effort to extend this type of analysis to
account for the effects of chemical reactions on the RT Boussinesq
turbulence will be described in a forthcoming article~\cite{09CLV}.

\begin{acknowledgments}

We wish to thank P.~Fischer for the permission to use the Nekton
code, A.~Obabko and P.~Fischer for the detailed help in using the
code, and J.R.~Ristorcelli for useful comments. This work was
supported by the U.S. Department of Energy at Los Alamos National
Laboratory under Contract No. DE-AC52-06NA25396 and under Grant No.\
B341495 to the Center for Astrophysical Thermonuclear Flashes at the
University of Chicago.

\end{acknowledgments}







\end{document}